\documentstyle[prb,aps,psfig]{revtex}
\begin{document}
\preprint{Phys. Rev. B}
\draft
\twocolumn[\hsize\textwidth\columnwidth\hsize\csname @twocolumnfalse\endcsname
\title{Monte Carlo study of the superfluid weight in doped antiferromagnets}
\author{Gregory C. Psaltakis}
\address{Department of Physics, University of Crete, and Research Center 
         of Crete, Heraklion, GR-71003, Greece}
\date{Received \today}
\maketitle
\begin{abstract}
The phase fluctuations of the condensate in doped antiferromagnets, described 
by a $t$-$t^{\prime}$-$J$ model and a suitable $1/N$ expansion, provide a 
mechanism for a Kosterlitz-Thouless (KT) type of transition to a superconducting
state below $T_{c}$. In this paper, we present a Monte Carlo study of the 
corresponding superfluid weight $D_{s}(T)$ in the classical (large-$N$) limit,
as a function of temperature and doping. Consistent with generic experimental 
trends, $D_{s}(T)$ exhibits a $T$-linear decrease at low temperatures, with the 
magnitude of the slope $D_{s}^{\prime}(0)$ increasing upon doping. Finite-size 
scaling in the underdoped regime predicts values for the dimensionless ratio 
$A=k_{B}T_{c}/D_{s}(0)$ of order unity, with $A=0.4435(5)$ in the 
half-filled-band limit, thus confirming $D_{s}(0)$ as the fundamental energy 
scale determining $T_{c}$. Our Monte Carlo results for $D_{s}(T)/D_{s}(0)$ vs 
$k_{B}T/D_{s}(0)$, at 10\% hole doping, are found to be in reasonable agreement
with recent measurements on La$_{2-x}$Sr$_{x}$CuO$_{4}$, with $x=0.10$, 
throughout the temperature range below the theoretical KT transition 
temperature $T_{c}$.
\end{abstract}
\pacs{PACS numbers: 74.40.+k, 74.20.Mn, 74.25.Bt}
\vskip2pc]

\narrowtext
\section{Introduction}
\label{sec:intro}
The quest on the nature of the superconducting transition observed in doped 
antiferromagnets, such as the lamellar high-$T_{c}$ copper-oxides, has entered 
a new era. Indeed, a consensus is emerging\cite{Millis99} on the importance of 
the zero-temperature superfluid weight $D_{s}(0)$ as the fundamental energy 
scale determining the corresponding transition temperature $T_{c}$. An early 
clue for this result came from the empirical Uemura relation,\cite{Uemura89-93}
valid for underdoped cuprates, which shows roughly a proportionality between 
the independently measurable quantities $T_{c}$ and $D_{s}(0)$, 
\begin{equation}
k_{B}T_{c}=AD_{s}(0) \;,
\label{eq:Tc}
\end{equation}
where $A$ is a dimensionless parameter of order unity. Given that the superfluid
weight $D_{s}(T)$, at a general temperature $T$, measures the 
``phase stiffness'' of the condensate, i.e., the energy cost to produce spatial
variations of its phase, Emery and Kivelson\cite{Emery95} argued that 
(\ref{eq:Tc}) provides strong evidence for a phase-fluctuation driven 
superconducting transition. Of course, in strictly two-dimensional systems with 
continuous U(1) gauge (phase) symmetry, melting of long-range phase coherence 
with increasing temperature, and hence loss of superfluidity of the charge 
carriers, proceeds by thermal generation and subsequent unbinding of 
vortex-antivortex pairs. The detailed description of this transition is given
by the Kosterlitz-Thouless (KT) theory\cite{KT73,KT-reviews} in the context of
the classical XY model. A characteristic feature of the KT transition is the 
discontinuous drop of $D_{s}(T)$ to zero at $T=T_{c}$, as $T_{c}$ is 
approached from below.\cite{Nelson77} Despite the inevitable rounding of this 
discontinuity by the weak coupling between the copper-oxide layers, 
Corson {\it et al}.,\cite{Corson99} in a remarkable experiment, have provided 
direct evidence of the KT nature of the superconducting transition by 
observing dynamic effects of thermally generated vortices in the 
frequency-dependent conductivity of underdoped cuprate thin films.

A common feature of theoretical models\cite{Roddick95,Coffey95,Carlson99}
assuming a phase-fluctuation mechanism for the superconducting transition, is 
the prediction of a $T$-linear decrease in $D_{s}(T)$ at temperatures well 
below $T_{c}$, in accordance with the experimental 
observations.\cite{Bonn96-95,Panagopoulos99} However, all the aforementioned 
models, do not incorporate explicitly the doping dependence of $D_{s}(T)$ and 
therefore provide no framework for a proper treatment of the high-$T_{c}$ 
cuprate superconductors as doped antiferromagnets. This is a major drawback 
since the proximity of the doped materials to the Mott insulating 
antiferromagnetic state at half-filling ($n_{e}=1$) leads\cite{Uemura89-93}
to a vanishing $D_{s}(0)$ and $T_{c}$, as the hole concentration $(1-n_{e})$ 
tends to zero. The importance of the antiferromagnetic fluctuations, even well 
within the underdoped regime, has been recently confirmed in a dramatic way. 
Specifically, the elastic neutron-scattering measurements\cite{Kimura99} 
in pristine La$_{2-x}$Sr$_{x}$CuO$_{4}$, with $x=0.10, 0.12$, have established
the presence of {\em static} long-range antiferromagnetic order in the 
superconducting ground state, suggested by earlier work.\cite{Suzuki98} 
This conclusion is supported also by the subsequent observation of conventional
two-magnon Raman scattering\cite{Lin00} in the same material.

The above discussion underlines the need for models of the high-$T_{c}$ cuprate
superconductors that can account, on the same footing, for the intimately 
coupled phase and spin fluctuations of the condensate. Such a model has been 
suggested some time ago\cite{Psaltakis93} and shown, by one of the authors, to 
display flux quantization and a finite superfluid weight, i.e., 
superconductivity, in its ground state.\cite{Psaltakis98} Our model consists of
a $t$-$t^{\prime}$-$J$ Hamiltonian and a suitable $1/N$ expansion that provide 
a framework for the study of: (i) the ground-state properties, using standard 
``spin-wave'' techniques which allow the incorporation of leading 
quantum-fluctuation effects, and (ii) the finite-temperature properties, using
an associated classical (large-$N$ limit) energy functional and corresponding 
partition function in terms of which important physical quantities can be 
readily expressed and calculated by Monte Carlo simulation. Point (i) was the 
subject of earlier works\cite{Psaltakis93,Psaltakis98} and the results, 
especially for the optical properties, were found to provide support for our 
effective model when compared with experiment. In the present paper we consider 
point (ii), focusing on the study of the superfluid weight as a function 
of temperature and doping, a subject of current experimental and theoretical 
interest.

In Sec.~\ref{sec:model} we give a brief description of our effective 
$t$-$t^{\prime}$-$J$ Hamiltonian and the associated classical energy functional
and partition function emerging in the large-$N$ limit. In Sec.~\ref{sec:D(T)} 
we derive an explicit expression for the superfluid weight $D_{s}(T)$ which we 
study by Monte Carlo simulation using the standard Metropolis 
algorithm.\cite{Binder97} We present results for the shape of the scaled 
curve, $D_{s}(T)/D_{s}(0)$ vs $k_{B}T/D_{s}(0)$, the value of the dimensionless
ratios $k_{B}T_{c}/D_{s}(0)$ and $D_{s}(T_{c})/k_{B}T_{c}$, as well as the 
value of the zero-temperature slope $D_{s}^{\prime}(0)$, clarifying their 
doping dependence in the regime of interest, i.e., close to half-filling. 
The KT nature of the transition to the superconducting state is supported by 
finite-size scaling analysis using the relevant Weber-Minnhagen scaling 
formula.\cite{Weber88} In particular, our numerical results for the temperature
dependence of $D_{s}(T)$, at 10\% hole doping, compare reasonably well with 
recent measurements\cite{Panagopoulos99} in underdoped 
La$_{2-x}$Sr$_{x}$CuO$_{4}$, for the corresponding concentration value 
$x=0.10$. Furthermore, the zero-temperature slope of the superfluid weight is 
predicted to approach, close to half-filling ($n_{e}\rightarrow 1$), the 
universal value: $D_{s}^{\prime}(0)=-k_{B}/2$, consistent with the available 
experimental data.\cite{Bonn96-95,Panagopoulos99} The latter limiting value is 
shown to be a hallmark of the strong antiferromagnetic correlations, present in
this temperature and doping regime. Our concluding remarks are summarized in 
Sec.~\ref{sec:conclusions}.

\section{Effective model}
\label{sec:model}
The effective model under consideration is described by a $t$-$t^{\prime}$-$J$ 
Hamiltonian expressed in terms of Hubbard operators 
$\chi^{ab}=|a\rangle\langle b|$ as
\begin{equation}
H=-\sum_{i,j}t_{ij}\chi^{0\mu}_{i}\chi^{\mu 0}_{j}
+{\textstyle\frac{1}{2}}J\sum_{\langle i,j\rangle}
(\chi^{\mu\nu}_{i}\chi^{\nu\mu}_{j}-\chi^{\mu\mu}_{i}\chi^{\nu\nu}_{j}) \;,
\label{eq:H-quantum}
\end{equation}
where the index 0 corresponds to a hole, the Greek indices $\mu,\nu,\ldots$ 
assume two distinct values, for a spin-up and a spin-down electron, and the 
summation convention is invoked. Here $J$ is the antiferromagnetic 
spin-exchange interaction between nearest-neighbor sites ${\langle i,j\rangle}$
on a square lattice endorsed with periodic boundary conditions and a total
number of sites $\Lambda=\Lambda_{x}\times\Lambda_{y}$, where 
$\Lambda_{x}=\Lambda_{y}$. For the hopping matrix elements $t_{ij}$ we assume
\begin{equation}
t_{ij}=\left\{\begin{array}{cl}
       t & \mbox{if $i,j$ are nearest neighbors} \\
       -t^{\prime} & \mbox{if $i,j$ are next nearest neighbors} \\
       0 & \mbox{otherwise} \;.
       \end{array}\right.
\label{eq:hopping}
\end{equation}
The conventions in (\ref{eq:hopping}) incorporate opposite signs for $t$ and 
$t^{\prime}$ as it is appropriate for the hole-doped cuprates.\cite{Hybertsen90}
In Ref.~\onlinecite{Psaltakis93} we generalized the local constraint associated
with (\ref{eq:H-quantum}) to $\chi^{00}_{i}+\chi^{\mu\mu}_{i}=N$, where $N$ 
is an arbitrary integer, and considered the commutation properties of the 
$\chi^{ab}$ operators to be those of the generators of the U(3) algebra. 
A Holstein-Primakoff realization for the latter algebra in terms of
hard-core bosons resolves then explicitly the local constraint and can be used 
to develop a perturbation theory based on the $1/N$ expansion, restoring the 
physical value $N=1$ at the end of the calculation.

In the presence of an external magnetic flux $\Phi$, threading the 
two-dimensional lattice in an Aharonov-Bohm torus geometry, the hopping matrix 
elements $t_{ij}$ are modified by the well known Peierls phase factor 
and should be substituted in (\ref{eq:H-quantum}) according to
\begin{equation}
t_{ij}\leadsto t_{ij}e^{iA_{ij}} \;, \;\;\;
\mbox{with} \;\; A_{ij}=\frac{2\pi\Phi}{\Lambda_{x}\Phi_{0}}
({\bf R}_{i}-{\bf R}_{j})\cdot{\bf e}_{x} \;.
\label{eq:Peierls}
\end{equation}
Here ${\bf R}_{i}$ is the position vector for lattice site $i$, ${\bf e}_{x}$ 
is the unit vector along the $x$-axis encircling the flux lines and 
$\Phi_{0}=2\pi\hbar c/q$ is the so-called flux quantum. As argued in 
Ref.~\onlinecite{Psaltakis98}, in the context of the present {\em effective} 
model, whereby carriers are treated as hard-core bosons, the charge $q$ 
entering $\Phi_{0}$ should be set equal to $q=2e$, where $e$ is the electronic 
charge. 

In the large-$N$ limit ``condensation'' occurs, i.e., the Bose operators become
classical commuting fields. Considering only uniform density configurations, 
the corresponding classical energy functional resulting from 
(\ref{eq:H-quantum})--(\ref{eq:Peierls}) takes the form\cite{Psaltakis98}
\begin{eqnarray}
H(\Phi) &=& -n_{e}(1-n_{e})\sum_{i,j}t_{ij} 
\nonumber \\
&& \left[\cos\frac{\theta_{i}}{2}\cos\frac{\theta_{j}}{2} 
\cos\left(A_{ij}+\frac{\psi_{i}-\psi_{j}-\phi_{i}+\phi_{j}}{2}\right)\right.
\nonumber \\
&& \left.+\sin\frac{\theta_{i}}{2}\sin\frac{\theta_{j}}{2} 
\cos\left(A_{ij}+\frac{\psi_{i}-\psi_{j}+\phi_{i}-\phi_{j}}{2}\right)\right]
\nonumber \\
&&+\frac{n_{e}^{2}}{4}J\sum_{\langle i,j\rangle}
[\cos\theta_{i}\cos\theta_{j}
+\sin\theta_{i}\sin\theta_{j}\cos(\phi_{i}-\phi_{j})-1] \;.
\nonumber \\
\label{eq:H-classical}
\end{eqnarray}
where $n_{e}$ is the average electronic density, the angles 
$(\theta_{i},\phi_{i})$ determine the local spin direction, and the remaining 
parameter $\psi_{i}$ determines the local phase of the condensate. The above 
functional form makes apparent the coupling between the phase and spin 
variables of the condensate through the kinetic energy term, proportional to 
$t_{ij}$.

The description of the finite-temperature classical theory is now completed 
using the energy functional (\ref{eq:H-classical}) to construct the 
corresponding partition function $Z(\Phi)$ and free energy per lattice site 
$F(\Phi)$,
\begin{equation}
Z(\Phi)=e^{-\beta\Lambda F(\Phi)}
=\int\left(\prod_{i}\sin\theta_{i}\,d\theta_{i}\,d\phi_{i}\,d\psi_{i}\right)
e^{-\beta H(\Phi)} \;,
\label{eq:partition-function}
\end{equation}
where $\beta=1/(k_{B}T)$ and the integrations at each lattice site $i$ extend 
over the intervals: $0\leq\theta_{i}\leq\pi$, $0\leq\phi_{i}\leq 2\pi$, and 
$0\leq\psi_{i}\leq 4\pi$. Invoking standard thermodynamic identities, important 
physical quantities can be readily expressed in terms of the partition function
(\ref{eq:partition-function}) and studied by Monte Carlo simulation. In the 
following we focus our discussion on the study of the superfluid weight 
$D_{s}(T)$.

\section{Superfluid weight}
\label{sec:D(T)}
At a finite temperature $T$, the superfluid weight (or helicity modulus) 
$D_{s}(T)$ is given by the curvature of the infinite lattice limit of the 
free energy $\Lambda F(\Phi)$ at $\Phi=0$,\cite{Yang62,Fisher73,Scalapino93-94}
\begin{equation}
D_{s}(T)=\Lambda\left(\frac{\Phi_{0}}{2\pi}\right)^{2}
\left[\frac{\partial^{2}F(\Phi)}{\partial\Phi^{2}}\right]_{\Phi=0} \;.
\label{eq:D(T):definition}
\end{equation}
Physically, $D_{s}(T)$ determines the ratio of the density of the superfluid 
charge carriers to their mass and hence can be related to the experimentally 
measurable in-plane magnetic (London) penetration depth, as noted later on in 
this section. Carrying out the second derivative with respect to $\Phi$ in 
(\ref{eq:D(T):definition}) we have more explicitly that
\begin{eqnarray}
D_{s}(T) &=& n_{e}(1-n_{e})\frac{2}{z\Lambda}\Bigg<\sum_{i,j}t_{ij}
|{\bf R}_{i}-{\bf R}_{j}|^{2}
\nonumber \\
&& \bigg[\cos\frac{\theta_{i}}{2}\cos\frac{\theta_{j}}{2} 
\cos\left(\frac{\psi_{i}-\psi_{j}-\phi_{i}+\phi_{j}}{2}\right)
\nonumber \\
&& +\sin\frac{\theta_{i}}{2}\sin\frac{\theta_{j}}{2} 
\cos\left(\frac{\psi_{i}-\psi_{j}+\phi_{i}-\phi_{j}}{2}\right)\bigg]\Bigg>
\nonumber \\
&& -[n_{e}(1-n_{e})]^{2}\,\frac{\beta}{\Lambda}\Bigg<\bigg\{\sum_{i,j}t_{ij}
[({\bf R}_{i}-{\bf R}_{j})\cdot{\bf e}_{x}]
\nonumber \\
&& \bigg[\cos\frac{\theta_{i}}{2}\cos\frac{\theta_{j}}{2} 
\sin\left(\frac{\psi_{i}-\psi_{j}-\phi_{i}+\phi_{j}}{2}\right)
\nonumber \\
&& +\sin\frac{\theta_{i}}{2}\sin\frac{\theta_{j}}{2} 
\sin\left(\frac{\psi_{i}-\psi_{j}+\phi_{i}-\phi_{j}}{2}\right)
\bigg]\bigg\}^{2}\Bigg> \;,
\nonumber \\
\label{eq:D(T)}
\end{eqnarray}
$z=4$ being the coordination number of the square lattice. As shown in 
Ref.~\onlinecite{Psaltakis93}, close to half-filling ($n_{e}\lesssim 1$) and 
for a sufficiently large $t^{\prime}$, the ground state of 
(\ref{eq:H-classical}), in the absence of magnetic flux ($\Phi=0$), is described
by a planar spin configuration ($\theta_{i}=\pi/2$) in which the local twist 
angles and phases exhibit long-range order according to: 
$\phi_{i}={\bf Q}\cdot{\bf R}_{i}$, 
$\psi_{i}={\bf Q}^{\prime}\cdot{\bf R}_{i}$, 
where ${\bf Q}=(\pi,\pi)$ is the usual spin-modulating antiferromagnetic 
wavevector and ${\bf Q}^{\prime}=(\pi,-\pi)$ is a phase-modulating wavevector.
The zero-temperature value of the superfluid weight follows then easily from 
(\ref{eq:D(T)}) as
\begin{equation}
D_{s}(0)=4t^{\prime}n_{e}(1-n_{e}) \;.
\label{eq:D(0)}
\end{equation} 
For the typical two-dimensional model with continuous symmetry under 
consideration, we expect that at low but finite temperatures, the long-range 
order will be destroyed by the proliferation of excited Goldstone modes, 
leading to a $T$-linear decrease of $D_{s}(T)$. At higher temperatures 
we expect that the thermal generation and subsequent unbinding of 
vortex-antivortex pairs will lead eventually to a discontinuous drop of
$D_{s}(T)$ to zero, at a critical point $T=T_{c}$, in a KT type of transition. 
In order to study numerically $D_{s}(T)$ in the whole temperature range and 
affirm the aforementioned physical picture, we performed a Monte Carlo 
simulation using the standard Metropolis algorithm.\cite{Binder97} 
Our calculations were carried out on small lattices, with typical sizes 
$\Lambda_{x}=8,16,32,64$. For a given temperature we performed of the order of 
$10^{4}$ thermalization steps and of the order of $10^{6}$ measurements. 
We considered values for the dimensionless ratios 
$\varepsilon=t^{\prime}/t=0.45$ and $t/J=1.0$, which are thought to be relevant
for the copper-oxide layers, and restricted our study to the underdoped regime,
i.e., to small $(1-n_{e})$ values up to 10\% hole doping. The latter 
restriction is dictated by the fact that models of the $t$-$t^{\prime}$-$J$ 
kind, being rather simple extensions of the Mott-Heisenberg antiferromagnetic 
insulator, cannot properly account for the nontrivial evolution of the 
electronic structure of the cuprates that occurs at higher doping values, 
namely, the closing of the pseudogap\cite{Timusk99} in the optimally doped and 
overdoped (Fermi liquid) regime.

\begin{figure}[h]
\centerline{\psfig{figure=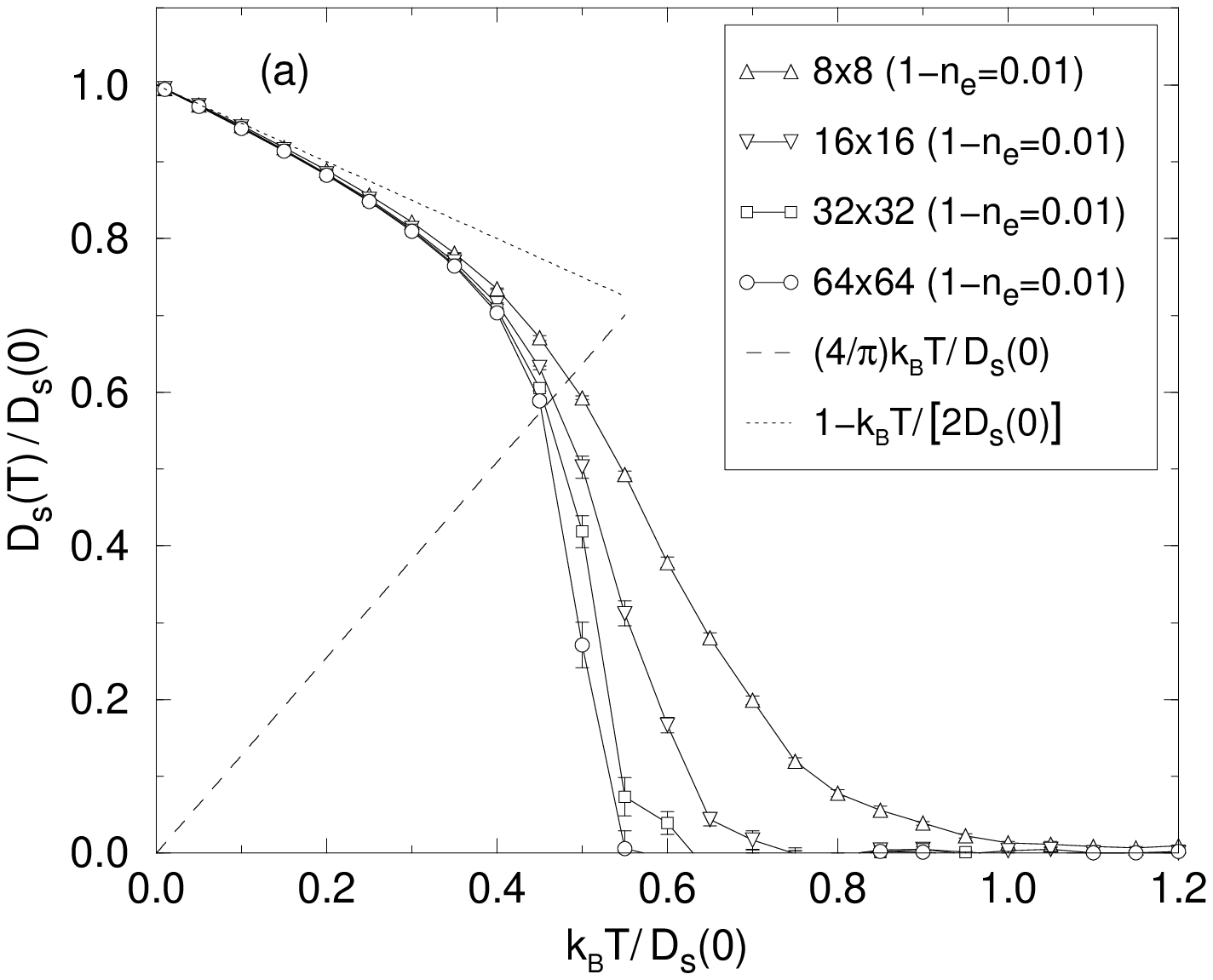,width=8.0cm}}
\centerline{\psfig{figure=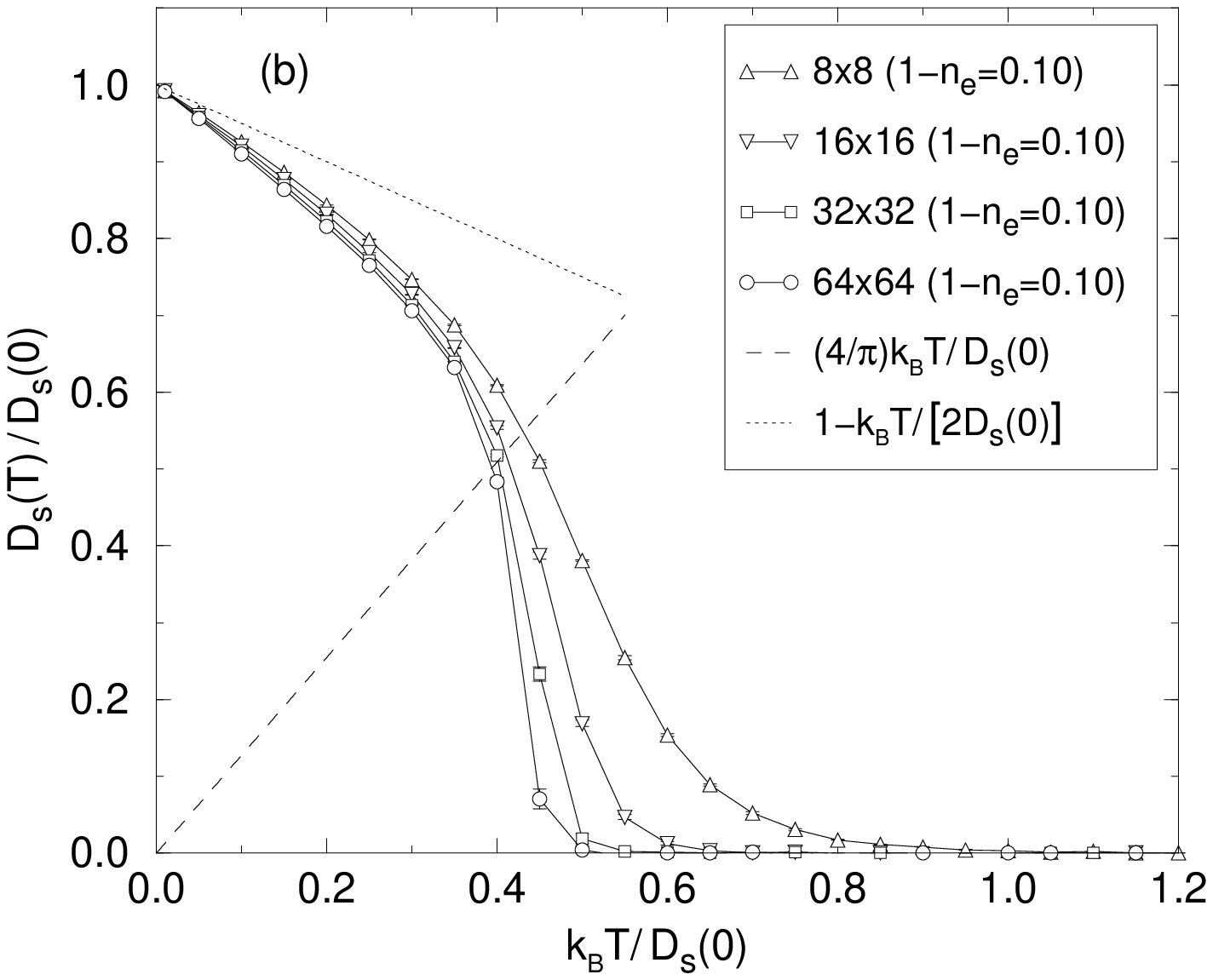,width=8.0cm}}
\caption{\label{fig:D(T)} Superfluid weight $D_{s}(T)$ vs temperature, for 
various lattice sizes, $\varepsilon=0.45$, $t/J=1.0$, and (a) $1-n_{e}=0.01$
[estimated $k_{B}T_{c}/D_{s}(0)=0.4346(9)$], (b) $1-n_{e}=0.10$ 
[estimated $k_{B}T_{c}/D_{s}(0)=0.3502(8)$]. Monte Carlo points above the 
corresponding estimated KT transition temperature $T_{c}$ are nonzero due to 
finite-size effects. Error bars are included but in most cases are smaller than
symbol size.}
\end{figure}

Typical Monte Carlo results for the superfluid weight vs temperature are shown
in Fig.~\ref{fig:D(T)}(a), for $(1-n_{e})=0.01$, and Fig.~\ref{fig:D(T)}(b), 
for $(1-n_{e})=0.10$. At low temperatures the superfluid weight has a weak 
finite-size dependence and displays the expected $T$-linear decrease. 
In particular, for $T\rightarrow 0$ and $n_{e}\rightarrow 1$, we have 
established the asymptotic form 
\begin{equation}
\frac{D_{s}(T)}{D_{s}(0)}=1-\frac{k_{B}T}{2D_{s}(0)} \;\;\; 
\mbox{for} \;\; T\rightarrow 0 \;\;\;
\mbox{and} \;\; n_{e}\rightarrow 1 \;. 
\label{eq:D(T)-low-T}
\end{equation}
The analytic expression (\ref{eq:D(T)-low-T}) is shown in Fig.~\ref{fig:D(T)}
by a dotted line. Evidently, this asymptotic line is approached very closely 
from below by the low temperature numerical data already in the case of 
the 1\% hole doping; see Fig.~\ref{fig:D(T)}(a). From (\ref{eq:D(T)-low-T}) 
follows that the zero-temperature slope of the superfluid weight approaches, 
close to half-filling, the parameter-free universal value:
\begin{equation}
D_{s}^{\prime}(0)=-k_{B}/2=-0.043\,\mbox{meV}\,\mbox{K}^{-1} \;\;\;
\mbox{for} \;\; n_{e}\rightarrow 1 \;.
\label{eq:D(0)-derivative}
\end{equation}
The upper limiting value (\ref{eq:D(0)-derivative}) is a rather stringent 
prediction of our theory and seems, indeed, to be consistent with the available 
experimental data\cite{Bonn96-95,Panagopoulos99} in the high-$T_{c}$ cuprate
superconductors. A comparison of Fig.~\ref{fig:D(T)}(a) with 
Fig.~\ref{fig:D(T)}(b) reveals an increase in the magnitude of the slope 
$D_{s}^{\prime}(0)$ upon doping, a trend also consistent with 
experiment.\cite{Bonn96-95,Panagopoulos99}

\begin{figure}[h]
\centerline{\psfig{figure=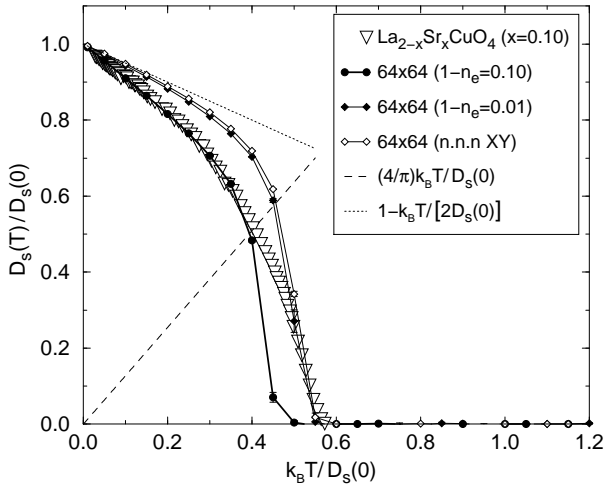,width=8.0cm}}
\caption{\label{fig:D(T)-experiment} Superfluid weight $D_{s}(T)$ vs 
temperature. Experimental data (triangles) on La$_{2-x}$Sr$_{x}$CuO$_{4}$, with
$x=0.10$, are extracted from the measurements of 
Panagopoulos {\it et al}.\protect\cite{Panagopoulos99} using 
Eq.~(\protect\ref{eq:D(T)-experiment}). Corresponding Monte Carlo results 
(circles) are calculated for lattice size $64\times 64$ and $\varepsilon=0.45$,
$t/J=1.0$, $1-n_{e}=0.10$. Also included are results for the same lattice size 
and $\varepsilon=0.45$, $t/J=1.0$, $1-n_{e}=0.01$ (filled diamonds), which 
should be compared with the results for the n.n.n XY model (opaque diamonds). 
The latter model corresponds to the limiting form of the present 
$t$-$t^{\prime}$-$J$ model, when $n_{e}\rightarrow 1$, as discussed in the 
text.}
\end{figure}

We emphasize that the asymptotic form (\ref{eq:D(T)-low-T}) is a physical 
consequence of the fact that close to half-filling the antiferromagnetic 
exchange energy, $Jn_{e}^{2}$, overwhelms the hole kinetic energy, 
$t_{ij}n_{e}(1-n_{e})$, and in particular the term $D_{s}(0)$ given by 
(\ref{eq:D(0)}). Hence, in the limit $n_{e}\rightarrow 1$ and in the relevant 
temperature range $k_{B}T\leq D_{s}(0)\ll Jn_{e}^{2}$ 
(so that $T\rightarrow 0$), the thermal average (\ref{eq:D(T)}) may be 
simplified by freezing the spin variables $(\theta_{i},\phi_{i})$ to their 
zero-temperature antiferromagnetic configuration and allowing fluctuations only
in the phase variables $\psi_{i}$. In this case, the vanishing overlap between 
the opposite sublattice spin states leaves the direct hopping $t^{\prime}$ 
between next-nearest-neighbor (n.n.n) sites as the only relevant process of 
charge transport. One can then easily show that the expressions for the energy 
functional (\ref{eq:H-classical}) and superfluid weight (\ref{eq:D(T)}) reduce 
to those of a classical XY model for the $\psi_{i}$ variables, but with only 
a n.n.n interaction $I$, where $I=D_{s}(0)/2$. The reduction of the structure 
of the phase fluctuations, close to half-filling, to that of the n.n.n XY model
and {\em not} to the commonly assumed\cite{Emery95,Psaltakis98,Alexandrov99} 
structure of the nearest-neighbor (n.n) XY model, is an important prediction 
and a direct consequence, in the context of our theory, of the presence of 
strong antiferromagnetic correlations in this regime. Numerically, the validity
of our argument becomes apparent in Fig.~\ref{fig:D(T)-experiment} displaying 
almost coinciding $D_{s}(T)$ Monte Carlo data for the n.n.n XY model 
(opaque diamonds) and the $t$-$t^{\prime}$-$J$ model with a very small hole 
concentration $(1-n_{e})=0.01$ (filled diamonds). Therefore, we may exploit the
detailed results of Appendix~\ref{app:xy-model} for the n.n.n XY model, see 
Eq.~(\ref{eq:xy-D(T)-low-T}), to conclude the asymptotic form 
(\ref{eq:D(T)-low-T}) and hence the limiting value (\ref{eq:D(0)-derivative}).
The latter value, $D_{s}^{\prime}(0)=-k_{B}/2$, being twice that of the 
n.n XY model (see Appendix~\ref{app:xy-model}), serves as a distinct hallmark 
of the sublattice structure of the strong antiferromagnetic correlations in the
limit $n_{e}\rightarrow 1$. Our observations here affirm also, by analogy to 
the well-known physics of the XY model, the presence of a KT transition for the 
superfluid weight of the $t$-$t^{\prime}$-$J$ model, when $n_{e}\rightarrow 1$,
and allow to transcribe relevant results for the former model, established in 
Appendix~\ref{app:xy-model}, to the latter, e.g.,
\begin{eqnarray}
&& \frac{D_{s}(T_{c})}{k_{B}T_{c}}=\frac{4}{\pi} \;\;\;
\mbox{for} \;\; n_{e}\rightarrow 1 \;, 
\label{eq:D(Tc)/kTc} \\
&& \nonumber \\
&& k_{B}T_{c}/D_{s}(0)=0.4435(5) \;\;\; 
\mbox{for} \;\; n_{e}\rightarrow 1 \;. 
\label{eq:kTc/D(0)}
\end{eqnarray}
In Fig.~\ref{fig:D(T)}  and Fig.~\ref{fig:D(T)-experiment} we present 
$(4/\pi)k_{B}T/D_{s}(0)$ by a short dashed line. According to 
(\ref{eq:D(Tc)/kTc}), in the limit $n_{e}\rightarrow 1$ the latter line should 
intersect the corresponding Monte Carlo data curve of $D_{s}(T)/D_{s}(0)$ vs 
$k_{B}T/D_{s}(0)$ precisely at the $k_{B}T_{c}/D_{s}(0)$ value given by 
(\ref{eq:kTc/D(0)}); see opaque (filled) diamonds in 
Fig.~\ref{fig:D(T)-experiment}.

It should be noted that in the actual simulations of the $t$-$t^{\prime}$-$J$ 
model we can only use finite, though possibly small, hole concentrations which 
inevitably lead to deviations from the limiting values 
(\ref{eq:D(Tc)/kTc})--(\ref{eq:kTc/D(0)}). In order to obtain rather accurately
the corresponding transition temperature $T_{c}$ we have used the finite-size 
scaling analysis of Weber and Minnhagen\cite{Weber88}, which is appropriate for
KT type of transitions. In this analysis one measures the chi-square values 
$\chi^{2}(T)$ of the fitting of the Monte Carlo data for the superfluid weight,
at each given temperature and for a sequence of small lattice sizes, to a 
certain scaling formula, derived by the latter authors from the Kosterlitz 
renormalization group equations.\cite{Kosterlitz74} Specifically, one assumes 
at each temperature $T$ the following $\Lambda_{x}$-dependence of the 
superfluid weight $D_{s}(T,\Lambda_{x})$,
\begin{equation}
\frac{\pi D_{s}(T,\Lambda_{x})}{2k_{B}T}
=R(T)\left(1+\frac{1}{2\ln[\Lambda_{x}/\Lambda_{0}(T)]}\right) \;,
\label{eq:scaling}
\end{equation}
where $\Lambda_{0}(T)$ is some characteristic length of the order of the 
lattice constant and has no singularity at $T=T_{c}$. The logarithmic lattice 
size dependence involved in (\ref{eq:scaling}) is characteristic of the 
presence of vortices in the KT transition. Strictly speaking (\ref{eq:scaling})
is correct only at the critical point $T=T_{c}$. Given now a value for 
$R(T_{c})$, the critical temperature can be determined by two alternative 
procedures:
(a) we fix $R(T)$ to be $R(T_{c})$ and use $\Lambda_{0}(T)$ as the only 
adjustable parameter in (\ref{eq:scaling}) to measure the chi-square values 
$\chi^{2}(T)$ of the fitting, in which case $T_{c}$ corresponds to the minimum
of the $\chi^{2}(T)$ curve, or 
(b) we use both $R(T)$ and $\Lambda_{0}(T)$ as adjustable parameters in 
(\ref{eq:scaling}) and determine $T_{c}$ from the point where the $R(T)$ 
curve crosses the line $R(T)=R(T_{c})$. The correct value for $R(T_{c})$ should
lead uniquely to the {\em same} value for $T_{c}$ in both procedures. 
The application of this finite-size scaling to the n.n.n XY model is summarized
pictorially in Fig.~\ref{fig:xy-scaling} of Appendix~\ref{app:xy-model} and 
justifies, as far as the $t$-$t^{\prime}$-$J$ model is concerned, the limiting 
values (\ref{eq:D(Tc)/kTc})--(\ref{eq:kTc/D(0)}) with $R(T_{c})\rightarrow 2$, 
for $n_{e}\rightarrow 1$.

For the finite doping value $(1-n_{e})=0.01$ or $(1-n_{e})=0.10$ 
(and fixed $\varepsilon=t^{\prime}/t=0.45$, $t/J=1.0$), the application of the 
aforementioned finite-size scaling analysis, using the lattice size sequence 
$\Lambda_{x}=4,6,8,10,12$, leads to: 
$k_{B}T_{c}/D_{s}(0)=0.4346(9)$ with $R(T_{c})=2.04$, or
$k_{B}T_{c}/D_{s}(0)=0.3502(8)$ with $R(T_{c})=2.51$, respectively.
To be sure, the presence in Fig.~\ref{fig:D(T)} of nonzero $D_{s}(T)$ values 
above the corresponding $T_{c}$, instead of a discontinuous drop to zero, is 
due to finite-size effects which grow rapidly above the estimated critical 
temperature; a typical behavior for a KT transition. Indeed, following
an original argument by Weber and Minnhagen,\cite{Weber88} we note that the 
success of the scaling formula (\ref{eq:scaling}) in the present model 
provides, {\it ipso facto}, strong evidence that the relevant phase transition
is of the KT type. Our results show a modest increase of the jump ratio 
$D_{s}(T_{c})/(k_{B}T_{c})=2R(T_{c})/\pi$ upon doping. The latter behavior 
seems physically similar to that known in the literature of the frustrated 
XY model\cite{Teitel83,Lee94} given that, in the context of the 
$t$-$t^{\prime}$-$J$ model, doping induces a form of dynamic frustration for 
the phase variables, via their inevitable coupling to the fluctuating spin 
variables. 

Furthermore, our results show that the dimensionless parameter 
$A=k_{B}T_{c}/D_{s}(0)$, introduced in context of the empirical Uemura relation
(\ref{eq:Tc}), is not doping independent but decreases modestly upon 
doping, while away from half-filling it also depends on the couplings 
$\varepsilon$ and $t/J$. Nevertheless, for rough theoretical estimates of the 
KT transition temperature $T_{c}$ in terms of $D_{s}(0)$, in the underdoped 
regime, one may always use the universal limiting value $A=0.4435(5)$, for 
$n_{e}\rightarrow 1$, quoted in (\ref{eq:kTc/D(0)}). We remind that that the 
latter value is characteristic of the n.n.n XY model and equals to half 
the corresponding value of the n.n XY model (see Appendix~\ref{app:xy-model})
commonly employed to this end.\cite{Emery95,Psaltakis98,Alexandrov99}.
Note that the use of $A=0.4435(5)$ in conjunction with (\ref{eq:Tc}) brings 
earlier theoretical overestimations of the KT transition temperature for the 
copper-oxides layers,\cite{Psaltakis98,Alexandrov99} derived with 
$A\approx 0.9$, down to more reasonable values. In all cases, the present 
analysis confirms $D_{s}(0)$ as the fundamental energy scale determining 
$T_{c}$ in the underdoped regime.

Having discussed the generic trends of the superfluid weight as a function of 
temperature and doping, in the underdoped regime, it is instructive to provide
a more direct comparison of our theory with experiment. To this end we note 
that in the lamellar high-$T_{c}$ superconductors, the experimental value for 
the superfluid weight per copper-oxide plane, $D_{s}^{\rm(exp)}(T)$, can be 
extracted from the directly measurable\cite{Bonn96-95,Panagopoulos99} in-plane 
magnetic (London) penetration depth, $\lambda_{ab}(T)$, using the 
relation:\cite{Emery95,Carlson99,Alexandrov99}
\begin{equation}
D_{s}^{\rm(exp)}(T)=\frac{(\hbar c)^{2}d}{4\pi q^{2}\lambda_{ab}^{2}(T)} \;,
\label{eq:D(T)-experiment}
\end{equation}
where $d$ is the average distance between planes and we remind that $q=2e$. 
Using the experimental data of Panagopoulos {\it et al}.\cite{Panagopoulos99} 
for $\lambda_{ab}(T)$ on the underdoped La$_{2-x}$Sr$_{x}$CuO$_{4}$, 
with $x=0.10$, and the structural parameter\cite{Jorgensen87} 
$d=6.64\:\mbox{\AA}$, we depict in Fig.~\ref{fig:D(T)-experiment} by triangles
the corresponding experimental values (\ref{eq:D(T)-experiment}) for the 
superfluid weight vs temperature. As a result of the weak coupling between the 
copper-oxide layers, the experimental data display no KT discontinuity but 
rather a continuous drop of the superfluid weight to zero, at a specific 
temperature value, that is not simply related to the ideal KT transition 
temperature of a copper-oxide monolayer. Corresponding Monte Carlo results for 
10\% hole doping are depicted in Fig.~\ref{fig:D(T)-experiment} by circles and 
calculated for a $64\times 64$ lattice, with $\varepsilon=0.45$ and $t/J=1.0$.
As noted earlier in this section, the theoretical KT transition temperature for
the latter set of parameters is $k_{B}T_{c}/D_{s}(0)=0.3502(8)$, while nonzero 
Monte Carlo points above this value are due to finite-size effects. Evidently,
our theoretical results (circles) in Fig.~\ref{fig:D(T)-experiment} compare 
reasonably well with the experimental data (triangles) throughout their common 
relevant temperature range, i.e., up to $T_{c}$. 

In Fig.~\ref{fig:D(T)-experiment} we also depict Monte Carlo results for 
1\% hole doping (filled diamonds), as well as results for the n.n.n XY model 
(opaque diamonds), thus providing the theoretical lineshape of the superfluid 
weight vs temperature, in the limit $n_{e}\rightarrow 1$. Clearly, it will
be very interesting to have measurements of the in-plane magnetic penetration 
depth on La$_{2-x}$Sr$_{x}$CuO$_{4}$, with hole concentration $x$ as small as 
it is experimentally possible, to compare with the present definite theoretical 
prediction.

\section{Concluding remarks}
\label{sec:conclusions}
In this paper we have presented a study of the temperature and doping 
dependence of the superfluid weight $D_{s}(T)$ in doped antiferromagnets 
described by the $t$-$t^{\prime}$-$J$ model 
(\ref{eq:H-quantum})--(\ref{eq:hopping}). Using Monte Carlo simulations and 
finite-size scaling analysis we have demonstrated that the phase fluctuations 
of the condensate, emerging in an appropriate classical (large-$N$) limit, 
drive superconductivity via a Kosterlitz-Thouless type of transition. Our 
theoretical results reproduce important generic experimental trends of 
$D_{s}(T)$, observed in the underdoped high-$T_{c}$ cuprate superconductors. 
This includes the $T$-linear decrease of $D_{s}(T)$ at low temperatures and the 
increase of the magnitude of the slope $D_{s}^{\prime}(0)$ upon doping. 

In particular, the sublattice structure of the strong antiferromagnetic 
correlations in the half-filled-band limit was shown to dictate the lineshape 
of $D_{s}(T)/D_{s}(0)$ vs $k_{B}T/D_{s}(0)$, for $n_{e}\rightarrow 1$, to be 
identical to that of the n.n.n XY model. In order to check this definite 
theoretical prediction we have suggested measurements of the in-plane 
magnetic penetration depth in very lightly doped cuprates. Here we should add
that higher order $1/N$ corrections are expected to renormalize 
downwards\cite{Psaltakis98} the fundamental energy scale $D_{s}(0)$ but, 
nevertheless, leave the lineshape of the {\em scaled} curve $D_{s}(T)/D_{s}(0)$
vs $k_{B}T/D_{s}(0)$ essentially intact.

The present study shares some common features with earlier 
works\cite{Roddick95,Coffey95,Carlson99} invoking a phase-fluctuation mechanism
for the high-$T_{c}$ superconductivity. On the other hand, all these works 
including ours are radically different from phenomenological 
approaches\cite{Lee97,Xiang98} that implicate the thermally excited nodal 
quasiparticles in a $d$-wave BCS superconducting state for the reduction of 
$D_{s}(T)$ with increasing temperature. The weak-coupling BCS type of 
approaches, however, are undermined by the absence of normal state 
quasiparticle peaks\cite{Loeser97} near the Brillouin zone points $(0,\pm\pi)$ 
and $(\pm\pi,0)$ where superconductivity is supposed to originate. At present 
it is still difficult to discern experimentally whether the temperature and 
doping dependence of $D_{s}(T)$ is dominated by phase fluctuations or by nodal 
BCS-like quasiparticle excitations. However, recent experiments in cuprate thin
films have provided strong evidence for the inherent two-dimensional character 
of superconductivity\cite{Saito98} and for the KT nature of the superconducting 
transition\cite{Corson99}.

\section*{Acknowledgments}
I would like to thank C. Panagopoulos for providing the penetration depth data 
and X. Zotos and G. Varelogiannis for stimulating discussions.

\appendix
\section{Generalized XY model}
\label{app:xy-model}
In the main body of the paper we noted that the structure of the phase 
fluctuations of the $t$-$t^{\prime}$-$J$ model under study reduces, in the 
half-filled-band limit, to that of a classical XY model with only 
next-nearest-neighbor (n.n.n) interactions. In order to clarify the properties 
of the latter model, in juxtaposition to those of the more conventional 
nearest-neighbor (n.n) XY model, we consider briefly in this appendix the 
following Hamiltonian
\begin{equation}
H_{\rm XY}=-\frac{1}{2}\sum_{i,j}I_{ij}\cos(\psi_{i}-\psi_{j}) \;,
\label{eq:xy-H}
\end{equation}
assuming
\begin{equation}
I_{ij}=\left\{\begin{array}{cl}
       (1-\alpha)I & \mbox{if $i,j$ are nearest neighbors}   \\
       \alpha I & \mbox{if $i,j$ are next nearest neighbors} \\
       0 & \mbox{otherwise} \;,
       \end{array}\right.
\label{eq:xy-exchange}
\end{equation}
where $\alpha$ is a free parameter, with $0\leq\alpha\leq 1$, and $I>0$. 
At each lattice site $i$ the angle $\psi_{i}$ varies in the 
interval: $0\leq\psi_{i}\leq 2\pi$. Evidently, $\alpha=0$ corresponds to 
the n.n XY model, while $\alpha=1$ corresponds to the n.n.n XY model.

\begin{figure}[h]
\centerline{\psfig{figure=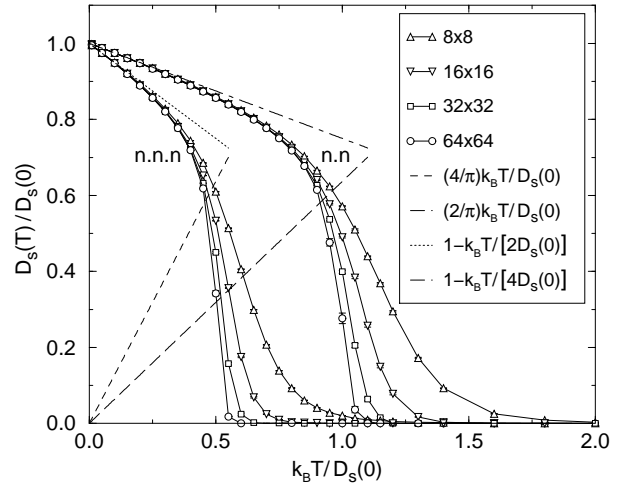,width=8.0cm}}
\caption{\label{fig:xy-D(T)} Superfluid weight $D_{s}(T)$ vs temperature for the
XY model with various lattice sizes and n.n.n or n.n interaction only. The 
crossings of the short (long) dashed line with the Monte Carlo data curves for 
the n.n.n (n.n) XY model provide estimates for the KT transition temperature 
$T_{c}$ of increasing accuracy, as the lattice size increases. Monte Carlo 
points above the corresponding estimated $T_{c}$ are nonzero due to finite-size
effects. The dotted (dash-dotted) line depicts the low-temperature asymptote for
the n.n.n (n.n) XY model, according to Eq.~(\protect\ref{eq:xy-D(T)-low-T}).}
\end{figure}

The superfluid weight (or helicity modulus) for the generalized XY model 
(\ref{eq:xy-H}) reads
\begin{eqnarray}
D_{s}(T) &=& \frac{2}{z\Lambda}\left\langle\frac{1}{2}\sum_{i,j}I_{ij}
|{\bf R}_{i}-{\bf R}_{j}|^{2}\cos(\psi_{i}-\psi_{j})\right\rangle
\nonumber \\
&& -\frac{\beta}{\Lambda}\left\langle\bigg\{\frac{1}{2}\sum_{i,j}I_{ij}
[({\bf R}_{i}-{\bf R}_{j})\cdot{\bf e}_{x}]\sin(\psi_{i}-\psi_{j})
\bigg\}^{2}\right\rangle \;,
\nonumber \\
\label{eq:xy-D(T)}
\end{eqnarray}
in agreement with corresponding early results.\cite{Teitel83} In view of 
(\ref{eq:xy-exchange}), the ground state configuration of (\ref{eq:xy-H}) is 
simply given by $\psi_{i}=0$, while the zero-temperature value of the 
superfluid weight follows immediately from (\ref{eq:xy-D(T)}) as 
\begin{equation}
D_{s}(0)=(1+\alpha)I \;.
\label{eq:xy-D(0)}
\end{equation}
Integrating the quadratic (Gaussian) fluctuations around the ground state 
configuration we obtain, after some lengthy algebra, the following 
low-temperature asymptotic expansion for the superfluid weight
\begin{equation}
\frac{D_{s}(T)}{D_{s}(0)}=1-[1+G(\alpha)]\frac{k_{B}T}{zD_{s}(0)} \;\;\; 
\mbox{for} \;\; T\rightarrow 0 \;.
\label{eq:xy-D(T)-low-T}
\end{equation}
Here $G(\alpha)$ is a dimensionless geometric factor given by
\begin{equation}
G(\alpha)=\frac{1}{\Lambda}\sum_{\bf q}\frac{\alpha(1-\delta_{\bf q})}
{(1-\alpha)(1-\gamma_{\bf q})+\alpha(1-\delta_{\bf q})} \;,
\label{eq:G(a)}
\end{equation}
with
\begin{equation}
\gamma_{\bf q}={\textstyle \frac{1}{2}}(\cos q_{x}+\cos q_{y}) \;, \;\;\;
\delta_{\bf q}=\cos q_{x}\cos q_{y} \;.
\end{equation}
We emphasize that $G(\alpha)$ is an increasing function of $\alpha$ with 
end-point values: $G(0)=0$ and $G(1)=1$. Hence, from (\ref{eq:xy-D(T)-low-T}) 
follows that the zero-temperature slope $D_{s}^{\prime}(0)$ evolves 
monotonically from $-k_{B}/4$ to $-k_{B}/2$, as the parameter $\alpha$ varies 
from 0 (n.n XY model) to 1 (n.n.n XY model); see dash-dotted and dotted lines
in Fig.~\ref{fig:xy-D(T)}.

Noting now that $|{\bf R}_{i}-{\bf R}_{j}|^{2}=2$, for n.n.n sites $i,j$, 
whereas $|{\bf R}_{i}-{\bf R}_{j}|^{2}=1$, for n.n sites $i,j$, a cautious 
inspection of (\ref{eq:xy-D(T)}) reveals that, in the thermodynamic limit and 
at each given temperature $T$, the superfluid weight of the n.n.n XY model 
should be twice as large the superfluid weight of the n.n XY model: 
$D_{s}^{(\alpha=1)}(T)=2D_{s}^{(\alpha=0)}(T)$. The latter property is explicit
in the low-temperature analytic results 
(\ref{eq:xy-D(0)})--(\ref{eq:xy-D(T)-low-T}) and we have confirmed its validity
in the whole temperature range by Monte Carlo simulation using the standard 
Metropolis algorithm with the parameters quoted in Sec.~\ref{sec:D(T)}. Indeed,
as shown in Fig.~\ref{fig:xy-D(T)}, the Monte Carlo data curves of 
$D_{s}(T)/D_{s}(0)$ vs $k_{B}T/D_{s}(0)$ for the n.n.n XY model coincide, 
within numerical error, with those for the n.n XY model, when the 
$k_{B}T/D_{s}(0)$ axis is scaled by a factor of 2. This agreement becomes 
better with increasing lattice size.

\begin{figure}[h]
\centerline{\psfig{figure=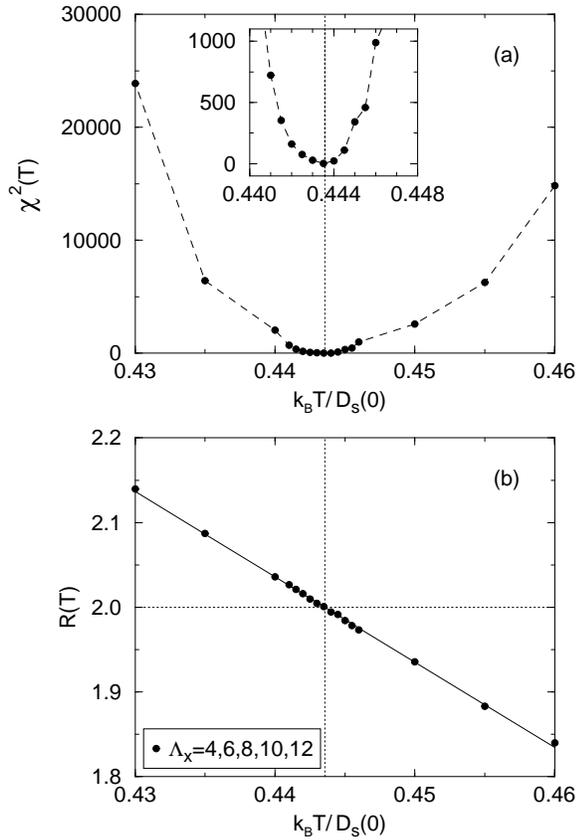,width=7.5cm}}
\caption{\label{fig:xy-scaling} Finite-size scaling around $T_{c}$ for the 
n.n.n XY model, according to Eq.~(\protect\ref{eq:scaling}) and for the lattice
size sequence $\Lambda_{x}=4,6,8,10,12$. (a) Chi-square values (circles) of the
fitting vs temperature with $R(T)$ fixed to 2. The inset is an enlarged view of
the curve around its minimum. (b) Coefficient $R(T)$ vs temperature. The dotted 
vertical line indicates the critical temperature at which the solid line 
crosses the dotted horizontal line [$R(T)=2$]. The solid line is determined by 
a linear fitting to the original data (circles). The estimated critical 
temperature is given by $k_{B}T_{c}/D_{s}(0)=0.4435(5)$.}
\end{figure}

For the conventional n.n XY model it is well-known\cite{Nelson77} that in the 
thermodynamic limit we have the jump ratio: $D_{s}(T_{c})/(k_{B}T_{c})=2/\pi$, 
or equivalently in the notation of (\ref{eq:scaling}), the value $R(T_{c})=1$.
Hence, in view of the simple relation 
$D_{s}^{(\alpha=1)}(T)=2D_{s}^{(\alpha=0)}(T)$, we anticipate for the 
n.n.n XY model the jump ratio: $D_{s}(T_{c})/(k_{B}T_{c})=4/\pi$, or 
equivalently, $R(T_{c})=2$. In order to demonstrate numerically the latter 
property we have carried out a finite-size scaling analysis for the 
n.n.n XY model based on the scaling formula (\ref{eq:scaling}), as described 
in detail Sec.~\ref{sec:D(T)}. The results are summarized in 
Fig.~\ref{fig:xy-scaling}. We note that the minimum of the $\chi^{2}(T)$ curve 
in Fig.~\ref{fig:xy-scaling}(a) occurs, within numerical error, at the 
{\em same} point were the $R(T)$ curve in Fig.~\ref{fig:xy-scaling}(b) crosses 
the line $R(T)=2$. Hence the assignment $R(T_{c})=2$ leads, indeed, to a 
uniquely determined value for $T_{c}$ which from the crossing point in 
Fig.~\ref{fig:xy-scaling}(b) is estimated to be 
$k_{B}T_{c}/D_{s}(0)=0.4435(5)$. The success of the finite-size scaling 
analysis validates then the assignment $R(T_{c})=2$ for the n.n.n XY model.

It is worth emphasizing that in terms of the dimensionless parameter 
$A=k_{B}T_{c}/D_{s}(0)$, the Monte Carlo results of the present appendix imply 
the value $A^{(\alpha=1)}=0.4435(5)$ for the n.n.n XY model and, of course, 
twice as large corresponding value for the n.n XY model, i.e., 
$A^{(\alpha=0)}=0.887(1)$. Note that the latter value for the n.n XY model 
agrees with the original estimate of Weber and Minnhagen\cite{Weber88} derived 
with the same finite-size scaling procedure. Pictorially, our results are 
manifest in Fig.~\ref{fig:xy-D(T)} were the crossings of the short (long) 
dashed line with the Monte Carlo data curves for the n.n.n (n.n) XY model 
provide estimates for the value of $A$ of increasing accuracy, as the lattice 
size increases.


\end{document}